\documentclass[noeprint,aps,prl,twocolumn,superscriptaddress,nolongbibliography]{revtex4-2}
\usepackage{amsmath}
\usepackage{graphicx}
\usepackage{amssymb}
\usepackage{enumitem}
\usepackage{mathrsfs}
\usepackage[bbgreekl]{mathbbol}
\usepackage{amsmath,amssymb,amsfonts,amsthm}
\usepackage{mathrsfs}
\usepackage[linktocpage=true]{hyperref}
\usepackage{upgreek} 
\usepackage{titlesec}

\usepackage{color}

\newcommand{\vortices}{{\mathfrak{c}}}
\newcommand{\st}{\sigma}
\newcommand{\ff}[1]{{\left\lfloor{#1}\right\rfloor}}
\newcommand{\reminder}[1]{{{\rm mod}\left({#1}\right)}}
\newcommand{\north}{{\rm N}}
\newcommand{\south}{{\rm S}}
\newcommand{\kcs}{{\rm k}}
\newcommand{\mN}{{p_+}}
\newcommand{\mS}{{p_-}}
\newcommand{\nN}{{n_+}}
\newcommand{\nS}{{n_-}}
\newcommand{\GG}{{\rm G}}
\newcommand{\GA}{{\mathfrak g}}

\def\nn{\nonumber}
\newcommand{\ii}{\mathrm{i}}
\newcommand{\ee}{\mathrm{e}}

\newcommand{\gflux}{\mathfrak{m}}
\newcommand{\ugot}{u}

\newcommand{\f}[2]{{\frac{#1}{#2}}}
\newcommand{\lc}{\varepsilon}
\newcommand{\p}[1]{{\left({#1}\right)}}
\newcommand{\comm}[1]{{\left[{#1}\right]}}

\newcommand{\spindle}{\mathbb{\Sigma}}
\newcommand{\Morb}{\mathbb{M}}

\newcommand{\Ft}{{\widetilde F}}

\newcommand{\dd}{{\rm d}}
\newcommand{\im}{{\rm i}}

\newcommand{\g}{\gamma}
\newcommand{\phit}{{\widetilde\phi}}
\newcommand{\la}{\lambda}
\newcommand{\lat}{{\widetilde\lambda}}

\newcommand{\z}{\zeta}

\newcommand{\zzt}{{v}}

\newcommand{\Ac}{{\mathcal A}}

\newcommand{\Fc}{{\mathcal F}}
\newcommand{\Gc}{{\mathcal G}}

\newcommand{\omeganot}{\omega}
\newcommand{\Cmat}{h}
\newcommand{\sigmavar}{\varphi}

\newcommand{\gammafug}{\gamma}
\newcommand{\pmin}{m_-} 
\newcommand{\pplu}{m_+}

\begin{document}

\title{The spindle index from localization}

\titlespacing*{\section}{0pt}{15pt}{15pt}
\titlespacing*{\subsection}{0pt}{15pt}{15pt}

\author{Matteo Inglese}
\author{Dario Martelli}
\author{Antonio Pittelli}
\email{apittelli88@gmail.com}
 \affiliation{Dipartimento di Matematica, 
 Universit\`a di Torino, 
Via Carlo Alberto 10, 10123 Torino, Italy}
\affiliation{INFN, Sezione di Torino,
 Via Pietro Giuria 1, 10125 Torino, Italy}

\begin{abstract}
\noindent  
We present a new supersymmetric index for three-dimensional ${\cal N}=2$ gauge theories defined on 
$\spindle \times S^1$, where $\spindle$ is a spindle, with twist or anti-twist for the  $R$-symmetry background gauge field.
We start examining  general supersymmetric backgrounds of
 Euclidean new minimal supergravity admitting two Killing spinors of opposite $R$-charges.
 We then focus  on $\spindle \times S^1$ and  demostrate how to  realise twist and anti-twist. We 
  compute the supersymmetric partition functions on such  backgrounds via  localization and show that these 
 are captured  by a  general formula, depending on the type of twist, 
which unifies and generalises  the superconformal and  topologically twisted indices.
\end{abstract}

\maketitle


\section{INTRODUCTION}\label{sec:intro}

Localization techniques \cite{Pestun:2007rz} are a tremendous  tool to calculate path-integrals of supersymmetric quantum field theories (SQFTs) on curved manifolds, giving access
to a profusion of  exact  results.
This letter focuses on three-dimensional  SQFTs, whereof two compelling   observables are the  partition functions on $S^2\times S^1$ endowed with an $R$-symmetry background gauge field,  with or without flux through the sphere, corresponding to the topologically twisted index \cite{Benini:2015noa} and the
 superconformal index  \cite{Imamura:2011su,Kapustin:2011jm}, respectively.
The large-$N$ limit of the former  provides a microscopic interpretation of the  entropy of magnetically charged supersymmetric black holes in AdS$_4$ \cite{Benini:2015eyy}.

The supersymmetric and \emph{accelerating} AdS$_4$ black hole presented in \cite{Ferrero:2020twa} 
displays a number of  striking  features, including a horizon with  orbifold singularities and supersymmetry preserved in a novel fashion.
 In the Euclidean  setting,
the conformal boundary  is  $\spindle\times S^1$,  where $\spindle=\mathbb{WCP}^1_{[\nN,\nS]}$ is a \emph{spindle}, \emph{i.e.} a weighted projective space  
specified  by two coprime positive integers $n_\pm$. The entropy  of this solution is reproduced by considering a family of Euclidean saddles extending the  black holes, where both  bulk and boundary metrics are complex \cite{Cassani:2021dwa}.

Motivated by these results, in this letter we  present  two partition functions of ${\cal N}=2$  Chern-Simons-matter theories defined on  $\spindle\times S^1$. 
Supersymmetry on $\spindle$ is preserved  by an  $R$-symmetry background gauge field $A$ satisfying only one \cite{Ferrero:2021etw} of the following conditions \footnote{We adopt the standard  convention  in which $A$ enters the KSE as in  (\ref{KSES}), whereas in  \cite{Ferrero:2021etw} $A_{\rm there} = 2 \, A_{\rm here}$.}:
\begin{align}
\label{twistantitwistdef}
 \int_{\spindle} \f{\dd A}{2\pi} = \frac{1}{2}\p{\frac{1}{\nS} +  \frac{\st}{\nN} } \equiv \frac{\chi_\st}{2} ~ , 
\end{align}
with $\st=\pm1$ being  configurations known as   \emph{twist} and \emph{anti-twist}, respectively. Here $\chi_+ =  \tfrac{1}{4\pi}\int_\spindle \sqrt{g}R$  is the orbifold Euler-characteristic of the spindle.  
We  employ the framework of rigid  new minimal supergravity and generalize   the analysis of  \cite{Closset:2012ru}  to accomodate 
 the  intrinsically complex backgrounds of \cite{Cassani:2021dwa}. We extend  the localization approach  to orbifolds and compute    partition functions  on  $\spindle\times S^1$  with both twist and anti-twist.   The technical details of the analysis will be spelled out in \cite{Inglese:2023tyc}.

\section{GENERAL COMPLEX BACKGROUNDS}
\label{sec:background}

 We consider  a general class of rigid supersymmetric backgrounds of Euclidean 
new minimal supergravity  preserving  two  Killing spinors $\z_\pm $ with  $R$-charges $\pm 1$, respectively. 
The Killing spinor equations (KSEs) are 
\begin{align}
\label{KSES}
&  (\nabla_\mu \mp \im A_\mu)\z_\pm = - \tfrac{H}{2} \g_\mu \z_\pm \mp \im V_\mu \z_\pm \mp  \lc_{\mu\nu\rho} \tfrac{ V^\nu}{2} \g^\rho \z_\pm ~ , 
  \end{align}
 where $A_\mu$ is the $R$-symmetry background gauge field, 
 $V_\mu$ is a globally defined co-closed one-form  and $H$ a scalar, a priori all complex-valued.   We emphasise  that $\z_\pm $ are not related by charge conjugation and, differently from previous literature,   the metric $g_{\mu\nu}$ can be complex-valued.

Following the conventions of  \cite{Closset:2012ru},  we introduce
\begin{align}
 \zzt = \z_+  \z_- ~ , \quad K^\mu = \z_+ \g^\mu \z_- ~ , \quad P_\pm^\mu =  \z_\pm \g^\mu \z_\pm / \zzt  ~ , 
\end{align}
where $\p{P_+^\mu}^* \neq P_-^\mu$. The KSEs imply that $K^\mu$ is a  complex Killing vector. Thanks to Fierz identities $\p{K, P_\pm}$ is  a canonical complex frame generating the line element 
\begin{align}\label{eq: metricinkpptframe}
\dd s^2 = \p{ K / \zzt }^2 - P_+ P_- ~ , 
\end{align}
with non-zero contractions being
\begin{align}
 \iota_K K = \zzt^2 ~ , \qquad  \iota_{P_-} P_+ = \iota_{P_+}  P_- = -2 ~ .
\end{align}
The KSEs determine the background fields $A_\mu,V_\mu,H$ in terms of $\p{K, P_\pm}$. 
 We are interested in complex backgrounds 
on a real three-dimensional space $\mathcal{M}_3$:  since $K$ is a complex Killing vector, it picks out two  angular coordinates $\sigmavar, \psi$, thus
\begin{align}
\label{Kvect}
K= \kappa \left(  \partial_ \psi+  \omeganot \partial_\sigmavar  \right) ~ , 
\end{align}
where $\kappa$ and $\omeganot$ are complex constants and $\sigmavar, \psi$ are $2\pi$-periodic. 
By rescaling $\zeta_\pm$ we  can set $\kappa=1$ without loss of generality. 
The most general metric on $\mathcal{M}_3$  invariant under  two real Killing vectors $ \partial_ \psi,\partial_\sigmavar$ can be written as 
     \begin{align}
\dd s^2  =  f^2 \dd x^2  + \Cmat_{ij} \dd\psi_i \dd \psi_j  \quad \mathrm{with}\quad \psi_1=\psi\, , \psi_2= \sigmavar\, , 
\label{mysimplemetric}
\end{align}  
          where the complex-valued functions $f(x)$ and  $\Cmat_{ij}(x)$, $i,j=1,2$ depend only on the coordinate $x$. In these coordinates we have 
          \begin{align}      
          K & =   (\Cmat_{11} +  \omeganot \Cmat_{12} )\dd\psi+  (\Cmat_{12} + \omeganot \Cmat_{22} )\dd\sigmavar \, ,  \nn\\
           P_\pm &  =  \mathrm{e}^{\pm2\im \theta} \big( \pm f  \dd x + \im (\sqrt{\Cmat} /\zzt )( -\omeganot \dd\psi+   \dd\sigmavar )\big) \, , \label{complexframe}\\
   \zzt^2 & =      \Cmat_{11} + 2 \omega \Cmat_{12}  +\omeganot^2  \Cmat_{22} \, , \nonumber
           \end{align}  
where $\Cmat=\mathrm{det}(\Cmat_{ij})$ and  $\theta \equiv \p{\alpha_1 \psi + \alpha_2 \sigmavar}/2$, with $\alpha_1,\alpha_2$ two real constants that we shall discuss momentarily.   
Defining   $A^C   \equiv A-\tfrac{3}{2}V$, the background fields read
\begin{align}
\label{niceAC}
V &  =  \frac{1}{v} \Big[\im H K  -  \star  \dd K \Big]  \, , \\
   A^C  & =   \f{v^3}{4f\sqrt{\Cmat} }\Big[ \frac{1}{\omeganot} \Big(\frac{\Cmat_{11}}{v^2}\Big)'\dd \psi  -\Big(\frac{\Cmat_{22}}{v^2}\Big)'\dd \sigmavar    \Big]+\dd \theta\, ,\nn
   \end{align} 
where  
a prime denotes derivative with respect to $x$. The function $H$ satisfies  $\mathcal L_K H = 0$ and is otherwise  arbitrary; however, it will enter in the localization computation
only  though the following combinations:
\begin{align}
 h_R & \equiv  \iota_K V - \im \zzt H    =  - \f1{2 \zzt} \star (K \wedge  \dd K) ~ , \nn\\
 \Phi_R & \equiv \iota_K\p{A^C +  V} - \im \zzt H =   ( \alpha_1 + \omeganot \alpha_2 )/2  ~ .
  \end{align} 
Taking  $\gamma^1 = \sigma^2, ~ \gamma^2 = \sigma^3, ~\gamma^3 = \sigma^1$, with $\sigma^i$ being the Pauli matrices,
in the   frame where    $e^1  = - f\, \dd x $ and
 \begin{align}\label{eq: orthonormalframe}
    e^2  = \sqrt{\tfrac{\Cmat}{ \Cmat_{11}}} \dd \sigmavar \,  ,   \quad e^3  =  \sqrt{\Cmat_{11}}\big(\dd \psi+ \tfrac{\Cmat_{12}}{\Cmat_{11}} \dd \sigmavar \big)\,  ,
\end{align}
  the Killing spinors satisfying (\ref{KSES}) take the form 
  \begin{align}
  & \zeta_{+ }   =  \ee^{\im \theta} \begin{pmatrix} u_1 , - u_2 \end{pmatrix}^T  ~ ,  & \zeta_{- }    = -   \ee^{- \im \theta} \begin{pmatrix}  u_2 ,  u_1 \end{pmatrix}^T  ~ ,
  \label{kilspins}
\end{align} 
where $T$  indicates transposition  and 
 \begin{align}
   u_{1,2}   & = 2^{-1/2}\sqrt{  v \mp   \omeganot \sqrt{{\Cmat}/{\Cmat_{11}}} }  ~ .
  \label{bestspinors}  
  \end{align}

\subsection{$\spindle\times S^1$ with twist and anti-twist}

The results presented so far  are purely local and apply to any space  $\mathcal{M}_3$,
 including e.g. $S^3$. We shall now restrict attention to $\mathcal{M}_3=\spindle \times S^1$.  We take  $\psi$  to parameterize  $S^1$ and $x \in [-1,1],\sigmavar$ as coordinates on $\spindle$.  This fixes the bevaviour of the functions  $f,\Cmat_{ij}$ near  the poles of $\spindle$, which  we denote as $\north\equiv\{x=1\}$ and $\south\equiv\{x=-1\}$.
By using reparameterization to set for simplicity $f=1$  and denoting by $\varrho_\pm$ the coordinates near  each pole, at leading order we  have
 \begin{align}
 \Cmat_{11}  \sim  \Cmat^\pm_{11}\, , \qquad \Cmat_{22}  \sim  \tfrac{1}{n_{\pm}^2} \varrho_\pm^2 \,    , \qquad \Cmat_{12} \sim \Cmat^\pm_{12} \varrho_\pm^2 \, , 
  \end{align}
at the north and south poles respectively, where $\Cmat^\pm_{12},    \Cmat^\pm_{11}$ are  complex  constants.  By using  (\ref{niceAC}) we   find  
\begin{align}
 \f1{2\pi} \int_{\spindle} \dd A & =  - \f12\p{  \frac{s_+}{\nN}+ \frac{s_-}{\nS} } ~ ,
\end{align}
where $s_\pm$ denote the signs of  the function $\zzt/ \sqrt{\Cmat_{11}}$ at the north and south poles, respectively.  Therefore, the type of supersymmetry-preserving twist is completely 
encoded in the behaviour of the function $\zzt$. 
Given a generic metric on $\spindle \times S^1$ and a parameter $\omeganot$, we   regard the third equation in  (\ref{complexframe}) as a definition of the function 
$\zzt$. From this, it   follows that generically the function  $\zzt/ \sqrt{\Cmat_{11}}$  has the same sign at both poles,  corresponding to the twist case. Instead,  the anti-twist is realized if the
 function  $\zzt/ \sqrt{\Cmat_{11}}$ has opposite sign at the poles. In this case the metric $\Cmat_{ij}$ and the parameter $\omeganot$ need be fine-tuned.

Requiring  $\zeta_\pm$ to be (anti-)periodic  under $\psi \sim \psi + 2\pi$ implies that 
  $\alpha_1 = n  \in \mathbb Z$.   On $\spindle$  the spinors 
  should be defined in two patches $\mathcal{U}_{\north/\south}$, with
  the non-singular gauge fields  related via gauge transformations, corresponding  to different  values of  $\alpha_2$. 
Specifically, the regular $A_{\north/\south}$ are obtained taking   $\alpha_2\!=\! s_+/\nN$ in  $\mathcal{U}_{\north}$ and     $\alpha_2\!= \! -s_-/\nS$ in  $\mathcal{U}_{\south}$.  
Finally, from (\ref{bestspinors})  one can see  that at the north and south poles of $\spindle$ the spinors behave as 
\begin{align}
  & \zeta_+^{\north/\south}   \sim   \begin{pmatrix} 1 , - s_\pm \end{pmatrix}^T  ~ ,  & \zeta_-^{\north/\south}     \sim\begin{pmatrix}  s_\pm , 1 \end{pmatrix}^T  ~ ,
\end{align}
so that indeed they have the same 2d-chirality for the twist  and the opposite 2d-chirality for  the  anti-twist   \cite{Ferrero:2021etw}.

By exploiting such formalism we can immediately jump at the localization computation of the partition functions on these backgrounds. 
Importantly,  these  will only depend  on $\omeganot$,  not on the specific representative metric, with the caveat explained above 
that for the anti-twist the metric  depends on $\omeganot$.
To illustrate these features, it is instructive to   consider the  explicit background
\begin{align}
\label{ourbesse}
\dd s^2 = f^2 \dd x^2 + \p{1-x^2}\p{\dd \sigmavar -  \Omega \dd \psi}^2 + \beta^2 \dd \psi^2 ~ ,
\end{align}
where $x\in [-1,1]$, $\sigmavar$ and $\psi$ have  $2\pi$-periodicities and  
\begin{align}
 f(x)  \sim n_\pm / \sqrt{2(1\mp x)}  ~ ,\qquad \mathrm{for}\quad x\to \pm 1 \, , 
\end{align}
so that  at any constant $\psi$, the coordinates $x,\sigmavar$ describe a spindle, equipped with a generic  metric. Although  we  choose $\beta$  real and positive for simplicity, the function $f(x)$ and the constants $\beta$ and $\Omega$ can be complex a priori.
The  frame $(K,P_\pm)$,  background fields $A,V,H$ and  Killing spinors $\z_\pm$ are then completely determined by our general formulae (\ref{complexframe}),
(\ref{niceAC}), (\ref{kilspins}) in terms of the metric functions $f(x),\Omega,\beta$ 
and the parameter $\omeganot$.

Let us briefly discuss how the two twists are realized for the family of metrics (\ref{ourbesse}). As $\zzt^2 =(1-x^2)(\omeganot-\Omega)^2+\beta^2$, if no relation is imposed between $\omeganot$, $\beta$ and $\Omega$,  then the  $R$-symmetry background field   realizes the   \emph{twist}.
As a special case, the standard topological twist corresponds to $\omeganot=\Omega$, yielding   $\zzt/\beta= - 1$,
 so that in (\ref{twistantitwistdef})  we have $\st=+1$.
Conversely, the \emph{anti-twist} is realized by choosing $\Omega=\omeganot \pm \im\beta$.
In this case we can take  $\zzt/ \beta=x$, 
so that  in (\ref{twistantitwistdef}) we have $\st=-1$.

\section{LOCALIZATION}

Let $\GG$ be a semi-simple Lie group with Lie algebra $\GA$, with $\mathfrak R_\GG$ a generic representation of $\GG$  and  ${\rm Ad}_\GG$ its adjoint representation. For a  three-dimensional  ${\cal  N}=2$ gauge theory,  the supersymmetry transformations of a vector multiplet $(\Ac , \upsigma, \la , \lat , D) \in {\rm Ad}_\GG $ 
and those of a chiral  multiplet  $(\phi, \psi, F) \in \mathfrak R_\GG$  can be  written as a cohomological complex \cite{Pestun:2007rz}. 
Solving the  BPS equations $\delta\psi=\delta \widetilde \psi=\delta\lambda=\delta\widetilde \lambda = 0$  yields the localizion locus of classical configurations
contributing by   $Z_{\rm class}$ to the partition function. Also, this formulation allows for recasting the computation of vector- and chiral-multiplets 1-loop determinants, $Z^{\rm VM}_{\text{1-L}}, Z^{\rm CM}_{\text{1-L}}$, as a cohomological problem. Below we will present the main steps of this procedure, referring to \cite{Inglese:2023tyc} for more details.
If $\GG = \GG_g \times \GG_f$,  the partition function of a  theory on $S^1 \times \spindle$ with gauge group $\GG_g$ and flavour group $\GG_f$ reads 
\begin{align}\label{eq: zs1spindlefull}
Z_{S^1 \times \spindle}\p{\omega,\ugot_f , \mathfrak f_f } =  \!\!\sum_{\mathfrak f_g \in \Gamma_{\mathfrak h_g} }\!\! \oint_{\mathcal C} \!\tfrac{\dd \ugot_g}{|W_g|} \,  \widehat Z \p{\ugot_g , \mathfrak f_g |\omega,\ugot_f , \mathfrak f_f }  ~ ,
\end{align}
where $\mathfrak h_g$ is the maximal Cartan-subalgebra of $\GA_g$,  $\Gamma_{\mathfrak h_g}$  the corresponding co-root lattice and $W_g$  its Weyl group; while $\ugot_{g,f}\!\! \in \!\!\mathfrak h_{g,f}$ and $ \mathfrak f_{g,f} \!\!\in \!\!\Gamma_{\mathfrak h_{g,f}} $ denote  gauge/flavour holonomies and fluxes, respectively;  $\widehat Z$ is the product of  $Z_{\rm class}$, $Z^{\rm VM}_{\text{1-L}}$ and $Z^{\rm CM}_{\text{1-L}}$ and
$\mathcal C$ is a suitable integration-contour for $\ugot_g$. The partition function  (\ref{eq: zs1spindlefull}) also depends on the spindle data $\p{n_\pm, \st}$, which we suppressed  to not clutter the notation. $Z_{S^1 \times \spindle}$ is related to the $\GG_f$-flavoured Witten-index of the   theory quantized on $\spindle$,   that is
\begin{align}
\label{traceformula}
I_{S^1 \times \spindle}
= {\rm Tr}_{\mathscr H\comm{\spindle}} \comm{\ee^{  - \upomega J - \upvarphi R  - \sum_i \upvarphi_i F_i} }  ~ ,
\end{align}
where $J, R,F_i$ generate   angular momentum, $R$-symmetry and flavour symmetries, respectively;  while  $\mathscr H\comm\spindle$ is the Hilbert space of states on the spindle, with either twist or anti-twist. We anticipate that the fugacities  $\upomega$ and $\upvarphi$ are not independent, but are related by 
\begin{align}
\label{lovelyconstraint}
\upvarphi - \frac{\upomega}{4} \chi_{-\st}& =   \im \pi n  \, , \qquad n \in \mathbb{Z} ~ .
\end{align}
For the anti-twist  ($\st=-1$), taking  $n=\pm 1$ so that the spinors are anti-periodic on $S^1$, this reproduces the relation found in \cite{Cassani:2021dwa} for the dual accelerating black holes.

\subsection{BPS locus and Chern-Simons term}

The  vector-multiplet BPS equations read
\begin{align}
  \iota_K (\star \Fc)    = - \zzt D -  \im h_R \upsigma \, , \qquad 
  \label{BPSvect12}
 \iota_K \Fc  = -   \im  \dd_\mathcal{A} ({ \zzt \upsigma} )  \, , 
\end{align}
where $\Fc$ is the field strenght of 
the gauge field $\Ac$. Neither solving  (\ref{BPSvect12}) nor computing $Z_{\rm class}$ or $Z_{\text{1-L}}$ require imposing reality conditions on fields.
 We  restrict for simplicity to  Abelian gauge fields, the non-Abelian generalization being straightforward. The BPS locus is parametrized by a gauge field  $\mathcal A$ obeying 
${\cal L}_K \mathcal A=0$, 
with  flux 
\begin{align}
\mathfrak f_G \equiv \frac{1}{2\pi}\int_{\spindle} \dd \mathcal A = \frac{\gflux}{\nN \nS} ~ , 
\end{align}
where  $\gflux\in \mathbb Z $~\cite{Ferrero:2021etw} and    we can always write 
 \begin{align}
& \gflux =  n_+ \pmin - n_- \pplu   ~ , \quad m_\pm \in \mathbb Z ~ . 
\end{align}
The second equation in
(\ref{BPSvect12}) is solved by 
\begin{align}
  \mathcal{A}_\psi +  \omeganot  \mathcal{A}_\sigmavar  -   \im \zzt \upsigma =  \Phi_G   ~ , 
  \label{fixsigma}
\end{align}
where $\Phi_\GG$ is an arbitrary  complex constant (in each  patch  $\mathcal{U}_{\north/\south}$).  The first equation in (\ref{BPSvect12}) yields the auxiliary field $D$.  The Abelian Chern-Simons term contributes to $Z_{\rm class}$ via  $Z_{\rm CS} = \ee^{- S_{\rm CS}}$, with
 \footnote{$\int \mathcal{A}\wedge \Fc $  is evaluated by extending $\mathcal{A}$ to a gauge field $\hat{\mathcal{A}}$ on $\mathcal{M}_4=$Disk$\times \spindle$ and computing $\int_{\mathcal{M}_4}\hat\Fc\wedge \hat \Fc$. } 
  \begin{align}
 \label{SCS}
S_{\rm CS} = \frac{\ii \, \kcs}{4\pi}\int \left(  \mathcal{A}\wedge \Fc  + 2 \ii  \star D \upsigma \right)   =  2\pi \im \,   \kcs \,  \mathfrak f_{\rm G} \ugot~ , 
 \end{align}
  evaluated on BPS configurations solving (\ref{BPSvect12}), where 
\begin{align}
\label{pippo}
\ugot & \equiv   \p{  \mathcal A_\psi^+ +  \mathcal A_\psi^-  - \im v^+\upsigma^+  -  \im v^-\upsigma^-  }/2~ , 
\end{align}
with the $\pm$ superscripts denoting  quantities evaluated at the \north/\south~ poles of $\spindle$. 
The differential operators acting on a chiral-multiplet field $\phi\in \mathfrak R_\GG$ 
with $R$-charge $q_R^\phi$ are 
\begin{align}
L_K & = \mathcal L_K - \im \, q_R^\phi \,  \Phi_R ~ , \nn \\
L_{P_\pm} & = \mathcal L_{P_\pm} - \im \,  q_R^\phi \, \iota_{P_\pm} \p{ A^C +  V } - \im  \, \iota_{P_\pm} \mathcal A       ~ ,
\end{align}
 and the chiral-multiplet BPS equations  read
 \begin{align}
  \delta^2 \phi = \p{  L_K + \Gc_{\Phi_\GG}  } \phi =0 ~ , \qquad   F + \im  L_{P_-}  \phi = 0 ~ ,    
  \label{BPSchiral}
 \end{align}
 where for a chiral-multiplet scalar field  $\phi \in \mathfrak R_\GG$ we have
\begin{align}
 \Gc_{\Phi_\GG} \phi = - \im  {\Phi_\GG} \circ_{\mathfrak R_\GG} \phi  ~ ,
\end{align}
where $\circ_{\mathfrak R_\GG}$ indicates the action of $\Phi_G$ according to the representation ${\mathfrak R_\GG}$. For arbitrary $q_R^\phi$ and  $\Phi_G$, regularity of the solutions to the first equation in (\ref{BPSchiral}) implies  $\phi=0$, while  $F=0$ follows from the second. 
Similarly, $\phit= \Ft=0$ for an anti-chiral multiplet.
Taking $\mathcal{A}$ and $\upsigma$ to be Hermitian, (\ref{fixsigma}) splits in two real equations determining $\mathcal{A}_\psi$ and $\upsigma$ in terms of $\mathcal{A}_\sigmavar$, which is constrained by the first equation in (\ref{BPSvect12}). We impose no reality conditions on the auxiliary  field $D$.

\subsection{1-loop determinants}

Using a set of cohomological  variables, the 1-loop determinant of a chiral multiplet of $R$-charge $r$ in $\mathfrak R_\GG$ can be cast in the form 
\begin{align}\label{eq: zchi1loopfuncdet}
Z^{\rm CM}_{\text{1-L}} = \f{\det_{ {\rm Ker}  L_{P_+} } \p{ L_K + \mathcal G_{\Phi_\GG} } }{\det_{ {\rm Ker}  L_{P_-} } \p{ L_K + \mathcal G_{\Phi_\GG} } } ~ ,
\end{align}
while the 1-loop determinant of a vector multiplet, $Z^{\rm VM}_{\text{1-L}}$, includes  the contribution of BRST-ghosts compatible with supersymmetry \cite{Pestun:2007rz}.
However, a standard argument  implies that formally  $Z^{\rm VM}_{\text{1-L}} = Z^{\rm CM}_{\text{1-L}}\p{r=2}|_{{\mathfrak R_\GG =\mathrm{Ad}_G}}$  \cite{Benini:2015noa}, so in the following we shall focus on chiral multiplets. 
 
The   modes $\mathscr F^\pm \in L^2\comm{\spindle \times S^1}$ contributing to  (\ref{eq: zchi1loopfuncdet}) obey the linear ODEs  $L_{P_\pm}{\mathscr F}^\pm=0 $ and are eigenfunctions of the operator $L_K + \mathcal G_{\Phi_\GG}$. The functional determinant of the latter is an infinite product of eigenvalues that,   after  regularization, explicitly provides (\ref{eq: zchi1loopfuncdet})   in terms of special functions \cite{Inglese:2023tyc}. In this letter
 we shall sketch an alternative  route  to the same result, namely  extracting the eigenvalues  from an \emph{equivariant orbifold  index theorem}.

In general $\GG  = \GG_g \times \GG_f$ and $\GA  = \GA_g \oplus \GA_f$, so that $\mathfrak R_{\GG} = \mathfrak R_g \otimes \mathfrak R_f $ and the chiral multiplet in $\mathfrak R_{\GG}$ is coupled to  dynamical vector-multiplets in ${\rm Ad}_{\GG_g}$  and background vector-multiplets  in ${\rm Ad}_{\GG_f}$, providing   flavour fugacities.  
For both the twist and the anti-twist the results are conveniently expressed via the following variables: 
\begin{align}\label{eq: z1loopfugacities}
\mN  & = \pplu- \st\,  r/2 ~ , \qquad \mS   = \pmin+ r/2 ~ , \nn\\
 	\mathfrak b  & = 1 + \st \ff{\st\, \mN/\nN} + \ff{-\mS/\nS} ~ , \nn\\
 \vortices  & =  \reminder{-\mS,\nS}/\nS -    \st\, \reminder{\st\,\mN,\nN} /\nN  ~ , \nn\\
 \gammafug   & = -    n/2  +   \omeganot \, \chi_{-\st}/4 ~ ,  \qquad q = \exp\p{ 2 \pi \im \, \omeganot} ~ , \nn\\
  y & =   q^{\vortices/2} \mathrm{e}^{ 2 \pi \im \p{ r \gammafug -  \ugot } } ~ ,
  \end{align}
 where $\ff{\mathrm{x}}$ is the floor of $\mathrm{x}$, namely the greatest integer less than or equal to $\mathrm{x}$; while $\reminder{ \mathrm{x},\mathrm{y}}$ is the reminder of the integer division of $\mathrm{x}$ by $\mathrm{y}$. Identifying $\omeganot=\im \upomega/(2\pi)$ and $\gamma =\im \upvarphi/(2\pi) $ reproduces  (\ref{lovelyconstraint}).  For a general $\GG$, one makes the replacements $m_\pm \to \rho\p{m_\pm} $ and $ \ugot \to \rho\p{\ugot} $, where $\rho = \rho_g + \rho_f$ is the weight  of $\mathfrak R_{\GG}$.

\subsection{Equivariant orbifold  index theorem}

 The 1-loop determinant on $\spindle \times S^1$ is obtained from   the equivariant   index of the   operator $L_{ P_-}$
 with respect to the  group action $g=\exp\p{- \im \epsilon \, \delta^2}$ with equivariant parameter  $\epsilon$  \cite{Pestun:2007rz}. By inspection,  $g = g_{\spindle} \, g_{S^1}$, with  $g_{S^1}\in U(1)$ acting   freely on $\spindle \times S^1$; then, the full index   $I^\st_{\spindle\times S^1}$  is the product of $I^\st_{\spindle}$ and the sum of irreducible characters of $U(1) $    \cite{Atiyah:1974obx}.

For the twist, $L_{ P_-}|_{\spindle} = \overline \partial_L$   at both poles of the spindle, where $\overline \partial_L$ is the Dolbeault operator  twisted by the holomorphic line orbibundle
 \begin{align}\label{eq: generalizedorbibundle}
   L= {\cal O}(n_- \mN - n_+ \mS)= {\cal O}(-\gflux- \tfrac{r}{2} (\nN+\st\nS))~,
 \end{align}
over $\spindle$, which requires $r \in 2 \, \mathbb Z$ for $L$ to be  well-defined. Notice that, if $\spindle = S^2$,  then $r\in \mathbb{Z}$ for the topological twist,  while $r$ is not quantized on the superconformal index background.  For the anti-twist, $L_{ P_-}|_{\spindle} =  \partial_L$ at the north pole and $L_{ P_-}|_{\spindle} = \overline \partial_L$ at the south pole of $\spindle$. The different behaviour of $L_{ P_-}$ for twist and anti-twist implies a different equivariant action on $T_{\rm N,S}\spindle$, making $\st$ appear in the final result.

In general, the index takes the form of  a sum of equivariant characteristic classes  of an \emph{associated orbifold} over the fixed-point set of a group action \cite{Vergne1996EQUIVARIANTIF}.  For the spindle, such a  set comprises  of $n_{+}$ copies of the north pole and $n_{-}$ copies of the south poles, giving 
\begin{align}
\label{rowindex}
I^\st_{\spindle}= \frac{1}{n_+} \sum_{j=0}^{n_+-1} \tfrac{\omega_+^{ - j \mN } q_+^{-\mN}}{ 1-\omega_+^{j \st }q_+^{\st}}  +  \frac{1}{n_-} \sum_{j=0}^{n_--1} \tfrac{\omega_-^{ - j \mS } q_-^{-\mS}}{ 1-\omega_-^{-j}q_-^{-1}}  ~ ,
\end{align}
where $q_\pm  = q^{1/n_\pm}$ and  $\omega_\pm = \ee^{2 \pi \im /n_\pm}$.
The parameters $q_\pm$ encode the linear $U(1)$ action near  the north and south-pole of $\spindle$, respectively, with the denominators arising from the action on the complexified tangent space and
 the numerators corresponding to the equivariant Chern characters of the line bundle $L$.

By recalling that  $r \in 2 \, \mathbb Z$, we can resum (\ref{rowindex})  into 
\begin{align}\label{eq: generalizedspindleindex}
  I^\st_{\spindle}    & =    \frac{  q^{-{\st}\ff{ {\st} \, \mN/n_+ }}}{1 - q^{\st}} + \frac{  q^{ \ff{-\mS/n_-} }}{1- q^{-1} } ~ ,
       \end{align}
{which is valid for both types of twists, simply choosing the sign of $\st$. It is interesting to discuss the case $\sigma=+1$, which is perhaps more familiar in the mathematics literature. In this case
the fraction in (\ref{eq: generalizedspindleindex}) simplifies in a polynomial, that can be written as
   \begin{align}
 I^{+1}_{\spindle } =  \left\{
\begin{array}{lr}
 q^{- \ff{ \mN/\nN}}  + \dots + q^{\ff{- \mS/\nS}}   ~ , \\
 -    q^{- \ff{\p{\mS-1}/\nS}} - \dots - q^{\ff{- \p{\mN+1}/\nN}}  ~ ,
\end{array}
\right.
\label{countsectionstwist}
  \end{align} 
  with  first and second line holding for ${\mathfrak b\geq0}$ and ${\mathfrak b\leq-1}$, respectively. 
 The expansions above match those of the equivariant index of $\overline \partial_L$, counting the $\mathfrak b $ holomorphic sections of $L$, 
and agree with the  Kawasaki-Riemann-Roch theorem \cite{Closset:2018ghr} in the non-equivariant limit: 
   \begin{align}
\lim_{q\to 1} I^{+1}_{\spindle } = \mathfrak b = \text{deg}(L)+1 ~ .
 \end{align} 
Including the contribution  of $g_{S^1}$ and fugacities for flavour and gauge symmetries  yields  
\begin{align}\label{eq: generalizedspindles1index}
  I^\st_{\spindle \times S^1}     & = \sum_{k \in \mathbb Z} \ee^{ - 2 \im k \epsilon } y^{-1}\p{ \frac{q^{\p{1-\mathfrak b}/2}}{1-q^\st} - \frac{q^{\p{1 + \mathfrak b }/2}}{1-q }  }   ~ ,
 \end{align}
with  $\mathfrak b,y, q$   reported in (\ref{eq: z1loopfugacities}).  
By Taylor-expanding in $q$ the two fractions in 
(\ref{eq: generalizedspindles1index}) we obtain an infinite set of eigenvalues for each pole of the spindle. These can then be converted 
into infinite products in a standard way.  Remarkably, after  suitable regularization, the final result for the 1-loop determinant can be written as single formula valid for both twists:
\begin{align}
\label{1loopgentwist}
   Z_{\text{1-L}}^{\rm CM} & =   \p{- y}^{\frac{1-\st - 2 \mathfrak b}{4}} q^{\frac{\p{1-\st}\p{\mathfrak b-1}}{8}} \frac{\p{ q^{\frac{1}{2}\p{1+\mathfrak b} } y^{-1} ; q  }_\infty }{\p{ q^{\frac{\st}{2}\p{1-\mathfrak b } } y^{-\st} ; q  }_\infty } ~ ,
\end{align}
where $\p{z;q}_n$ is the $n$-th $q$-Pochhammer symbol. For the twist $\st=+1$ and   (\ref{1loopgentwist})  simplifies as  the finite product
\begin{align}
\label{1looptwist}
    Z_{\text{1-L}}^{\rm CM} & =   \p{-y}^{-\mathfrak b/2}\p{ q^{\frac{1}{2}\p{ 1 - \mathfrak b} }  y^{-1} ; q}_{\mathfrak b}^{-1} ~ ,
\end{align}
with the ${\mathfrak b}$ factors appearing in the $q$-Pochammer symbol above  precisely corresponding to the contributions of the  ${\mathfrak b}$ sections of the line bundle $L$, counted by (\ref{countsectionstwist}).
For either choice of twist, invariance under large gauge transformations, corresponding to integer  shifts of  $\ugot$,  induces a $1/2$-shift of the  Chern-Simons level $\kcs$.  After root decomposition, the  1-loop determinant of a  vector multiplet turns out to be independent of the $R$-symmetry twist and, up to a regularization-dependent sign, it reads
\begin{align}
\label{1loopVM}
    Z_{\text{1-L}}^{\rm VM} &  =  \prod_{\alpha>0  }  \prod_{I = \pm}   \p{ {\rm z}^{-\alpha/2}  -  q^{\frac{\alpha_+}{2 n_+} + \frac{\alpha_-}{2 n_-} - \ff{ \frac{\alpha_I}{ n_I} } } {\rm z}^{\alpha/2}   }^{\mu_I} \nn\\
     &  \times q^{\frac{1}{8} (\mu_{-}-\mu_{+})\tfrac{\alpha(\mathfrak m)}{n_+n_-}} ~ ,
\end{align}
where $\alpha_\pm  = \alpha\p{m_\pm} $, ${\rm z}^\alpha = \mathrm{e}^{2 \pi \im \alpha\p{u} } $, $\alpha$  is the  weight of the adjoint representation, while
 \begin{align}
\mu_I \equiv
\begin{cases}
1 \quad \mathrm{if} \quad  \alpha_I/n_I \in \mathbb{Z} \\
0 \quad \mathrm{otherwise}
\end{cases}\, . 
\end{align}

Although  the classical contribution (\ref{SCS}) only depends  on the total  gauge-field flux $\gflux$, 
 the 1-loop determinants depend a priori  on $m_\pm$  through $\mathfrak{b}$, $\vortices$, 
defined in (\ref{eq: z1loopfugacities}), and through  $\alpha_\pm$. Taking  $a_\pm \in \mathbb Z$ such that  $a_- n_+  - n_- a_+ = 1$, we can parameterise
$m_\pm = (a_\pm + t \, n_\pm)\gflux$, where $t\in \mathbb{Z}$. Since both $\mathfrak{b}$ and $\vortices$ are independent of $t$,  $Z_{\text{1-L}}^{\rm CM}$ 
depends only on $\gflux$; similarly, one can   check that $Z^{\rm VM}_{\text{1-L}}$ depends only on $\gflux$. Thus, in the complete partition function  (\ref{eq: zs1spindlefull})   we have to sum only over 
the gauge flux  $\gflux\in  \mathbb{Z}$.
 Finally,  taking $n_+=n_-=1$ reduces the full partition function to the topologically twisted index \cite{Benini:2015noa} and the superconformal index  \cite{Kapustin:2011jm}  upon setting $\st=\pm 1$, respectively.

\section{DISCUSSION}

In this letter we demonstrated that  three-dimensional  ${\cal N}=2$  SQFTs can be defined 
on $\spindle \times S^1$,  endowed with both types of supersymmetry-preserving twists and 
that  the corresponding  partition functions  give rise to two novel indices. 
These  can 
 be expressed by a single formula,  
generalizing  and unifying the   superconformal and  topologically twisted indices. We  therefore refer to this as to the \emph{spindle index}.
Many more details 
and applications 
will be discussed in \cite{Inglese:2023tyc}. 
The expression we found for the 1-loop determinant (\ref{1loopgentwist}) resembles the 
supersymmetric observables involving vortex defects  computed
 in \cite{Hosomichi:2017dbc}
   and it would be interesting to investigate further their relationship. 
   We anticipate that the large-$N$ limit  of the spindle index should  reproduce the entropy functions associated to the  supersymmetric and  
  accelerating  AdS$_4$  black holes  \cite{Ferrero:2021ovq,Ferrero:2020twa}.
More generally, it should reproduce the  entropy functions presented in \cite{Boido:2022iye},   valid for an extensive class of
 three-dimensional
 ${\cal N}=2$ theories with gravity duals.
We expect  that  gravitational blocks  \cite{Benini:2016rke}, whose gluing yields these entropy functions, 
should  arise in the large-$N$ limit of the single fixed-point contributions to the orbifold equivariant  index  discussed here.
 Our findings suggest that  some observables of SQFTs compactified  on spindles and other 
orbifolds $\Morb_p$   can be computed via localization. 
In particular,  it would be interesting  to compute  orbifold partition functions 
of SQFTs on  $M_{d-p}\times \Morb_p$ 
and to prove the large-$N$
 gravitational block formulas conjectured in \cite{Faedo:2021nub,Faedo:2022rqx}.

\section*{ACKNOWLEDGMENTS}
\noindent We thank C. Closset,  P. Ferrero, H. Kim, S. Murthy,  J. Sparks and A. Zaffaroni  for useful comments.
 DM thanks A. Zaffaroni for enjoyable collaboration on related topics.

\end{document}